\documentclass{article}
\usepackage{spconf,amsmath,graphicx,hyperref}
\usepackage{tikz}
\usepackage[precision=2, unit=mm]{lengthconvert}
\usepackage{float} 
\usepackage{pifont} 
\usepackage{url}
\usepackage{graphicx}    
\usepackage{multirow}    
\usepackage{makecell}    
\usepackage{rotating}    
\usepackage{booktabs}    
\usepackage{pifont}      
\usepackage{xcolor}      
\usepackage{xspace}
\usepackage{array}
\usepackage{float}
\usepackage{caption} 

\newcommand{\cmark}{\ding{51}} 
\newcommand{\xmark}{\ding{55}} 
\newcommand{\xmarkgray}{\textcolor{lightgray}{\xmark}\xspace}

\title{Mix2Morph: Learning Sound Morphing from Noisy Mixes}

\name{ Annie Chu$^{1,2}$, Hugo Flores-García$^{2}$, Oriol Nieto$^{1}$,
Justin Salamon$^{1}$, Bryan Pardo$^{2}$, Prem Seetharaman$^{1}$}

\address{$^{1}$ Adobe Research, San Francisco, USA \quad
         $^{2}$ Northwestern University, Evanston, USA \\
         {anniechu@u.northwestern.edu}}

\begin{document}
\usetikzlibrary{patterns}
\ninept
\topmargin=0mm

\maketitle
\urlstyle{same}

\begin{abstract}
We introduce Mix2Morph, a text-to-audio diffusion model fine-tuned to perform sound morphing without a dedicated dataset of morphs. By finetuning on noisy surrogate mixes at higher diffusion timesteps, Mix2Morph yields stable, perceptually coherent morphs that convincingly integrate qualities of both sources. We specifically target sound infusions, a practically and perceptually motivated subclass of morphing in which one sound acts as the dominant primary source, providing overall temporal and structural behavior, while a secondary sound is infused throughout, enriching its timbral and textural qualities. Objective evaluations and listening tests show that Mix2Morph outperforms prior baselines and produces high-quality sound infusions across diverse categories, representing a step toward more controllable and concept-driven tools for sound design.
\end{abstract}

\begin{figure*}[!h]
  \centering
  \includegraphics[width=\textwidth]{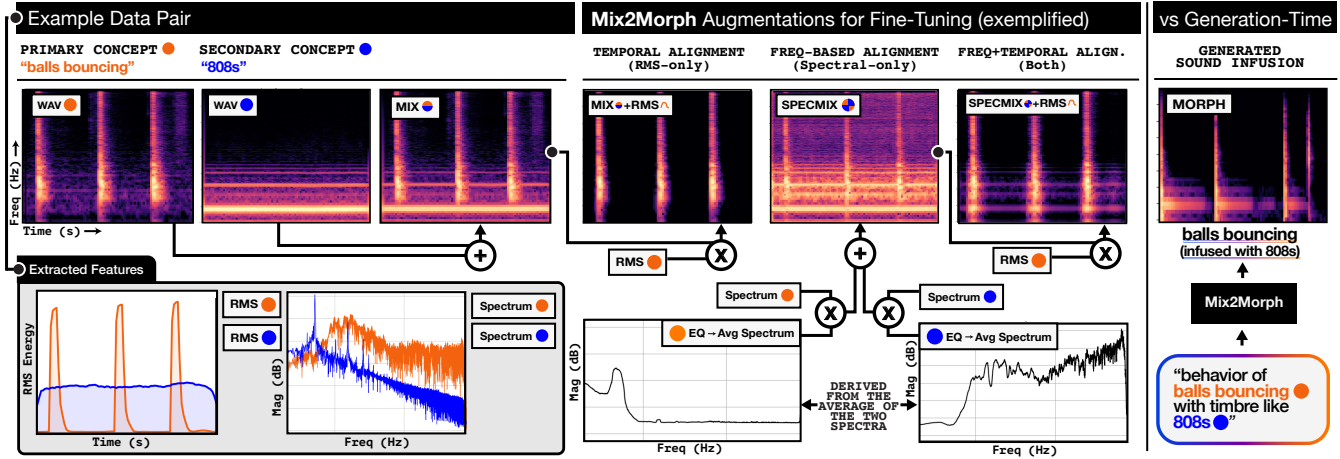}
  \caption{Mix2Morph pipeline for creating noisy surrogate morphs via augmented additive mixing of sound pairs, used as high-timestep training targets in our diffusion-based TTA model. This example (as well as others) can be listened to at  \textcolor{magenta}{\textbf{\url{https://anniejchu.github.io/mix2morph/}}}}.
  \label{fig:bigfig}
  \vspace{-1.0em}
\end{figure*}

\section{Introduction}
\label{sec:intro}
Sound design encompasses the creative process of shaping and combining audio to achieve expressive narrative goals. A common approach is \textit{sound morphing}: the fusion of two sounds into a single, coherent sound resembling both at once \cite{caetano2011morphing}. For instance, to design a menacing alien voice, a sound designer may combine animal vocalizations with human speech, producing a unified sound that is at once recognizable and alien. 

A central challenge in morphing is the generation of perceptually coherent midpoints: hybrids that convincingly sound like both sources at once. Human perception of sound is shaped primarily by two dimensions; temporal structure and spectral content, which provide natural anchors for coherence \cite{heller2022hybrid, oxenham2018we, smith2002chimaeric, Iverson1993IsolatingTDA}. When these anchors are not preserved, intermediate morphs become perceptually fragile: they may lose the identity of one or both sources, sound more like an additive mix than a fused sound, or drift into unnatural artifacts \cite{dixit2025learning, niu2024soundmorpher}. This issue is important in practice, as sound designers often aim not to fully transform one sound into another but to preserve the recognizable identity of a primary source while enriching it with qualities of a secondary source. To capture this practically motivated and perceptually distinctive case, we introduce the term \textit{sound infusion}. We define a sound infusion as a special case of asymmetric (or unbalanced \cite{dixit2025learning}) static morphing \cite{sethares2015kernel, caetano2012formal}, in which one sound acts as the primary source (e.g., a human voice) while another serves as the secondary (e.g., a lion's roar), contributing supplemental features throughout (e.g., a rough, roar-like texture woven into its timbre).

In this work, we focus on producing robust, high-quality, perceptually-coherent sound infusions where the temporal dynamics are rooted in the primary source while timbral qualities are jointly blended from both, as in \cite{tokui2025latent}. While sound infusion offers a practically motivated framing of the midpoint problem, prior work has focused on other morphing forms (e.g., dynamic or repetitive \cite{kazazis2016sound, sethares2015kernel, kamath2025morphfader}). Classic DSP methods, such as feature interpolation guided by perceptual descriptors \cite{caetano2011morphing, kazazis2016sound, caetano2019morphing} and cross-synthesis \cite{donahue2016extended, Puche2021CaesynthRT} generally produce good morphs when applied to pitched, harmonic sounds (e.g., instruments, vocals). However, they struggle with unpitched textures like ambient noise, environmental recordings, and everyday sound effects (SFX)--categories central to creative sound design. 

To broaden morphing beyond these limits, recent work has turned to deep learning \cite{gupta2023towards}.
Systems like MorphFader \cite{kamath2025morphfader} and SoundMorpher \cite{niu2024soundmorpher} extend the expressive scope of morphing, leveraging the pre-existing abilities of AudioLDM2 \cite{liu2024audioldm}, a text-to-audio (TTA) model. However, they frequently suffer from midpoint collapse: endpoint generations remain stable, but intermediate morphs are perceptually fragile and drift toward incoherent or additive-sounding mixes. 

A complementary line of work decomposes sounds into their dominant perceptual dimensions \cite{heller2022hybrid, oxenham2018we, smith2002chimaeric, Iverson1993IsolatingTDA}, temporal structure and spectral content, and blends them independently. Dixit et al. \cite{dixit2025learning} targeted temporal envelopes by interpolating amplitude contours via an autoencoder, while Tokui and Baker \cite{tokui2025latent}, building on granular resynthesis \cite{bitton2020neural, tralie2024concatenator}, used neural audio codecs to preserve a primary source’s temporal structure while infusing spectral content from a secondary source. This latter approach is especially relevant to sound infusions, but it remains fragile across diverse sound categories and often introduces audible, noise-like artifacts. 

A key bottleneck in training a model for morphing is the lack of training data exemplifying morphs, especially ones that both remain perceptually coherent across diverse sound categories and preserve the asymmetric dominance structure valued in creative sound design. In this work, we present Mix2Morph, a text-to-audio model designed for generating high-quality sound infusions. Mix2Morph's strength lies in a finetuning strategy that enables morph generation without requiring a dedicated morphing dataset. Inspired by the idea of ``making good data from bad data'' \cite{daras2025ambient}, we construct surrogate training data by repurposing noisy additive mixes, allowing the model to learn how to produce perceptually coherent morphs. Our contributions include:

\begin{itemize}
    \item Mix2Morph, a text-to-audio model for robust midpoint sound infusions, enabled by a finetuning strategy that repurposes pretrained generative models to perform morphing
    \item An automated method for constructing a surrogate training dataset for morphing, enabling model training without requiring a pre-existing morph corpus
    \item Extended objective metrics for evaluating morphing quality, including a \textit{directionality} measure and a proxy for morph identification via latent compressibility
    \item Objective and subjective validation that Mix2Morph outperforms both the prior base model and state-of-the-art deep learning baselines
\end{itemize}

\section{Proposed Methodology}
\label{sec:methods}
Training sound morphing models is hindered by the lack of high-quality morph datasets. We address this by constructing noisy surrogates, specially designed \textit{mixes} of two sounds, as training signals.

We adapt a ``no-waste'' dataset training strategy \cite{daras2025ambient}, where low-quality data are not discarded but instead assigned to higher diffusion timesteps. In diffusion training, the denoising objective depends on the noise level (timestep $t$): at low $t$, the model emphasizes recovering fine-grained details, while at high $t$, it focuses on coarse, global structure. We hypothesize that assigning noisy surrogate morphs to higher timesteps encourages the model to capture high-level morphing concepts while suppressing low-level mixing artifacts, thereby leveraging imperfect data without overfitting to artifacts. 

In our context, the low-quality data are \textit{simple additive mixes} of training pairs, which we treat as ``bad'' morphs. To strengthen their role as surrogate morphs, we apply temporal and spectral augmentations that encourage overlap and alignment, effectively bringing both sources into a shared structural and/or spectral space (Fig. ~\ref{fig:bigfig}). These noisy surrogates are used to train a model only as high-timestep training targets, never as  low-timestep  targets. At inference, we rely on refinement learned in pretraining to construct fine detail in low-$t$ timesteps. Crucially, this lets us construct a scalable training set without requiring high-quality morph examples, making this the first work to leverage noisy surrogate morphs for training at scale. 

\subsection{Augmentation Techniques}
\textbf{\textit{Temporal Alignment (RMS Anchoring)}:} We begin with an additive mix of a primary and secondary waveform, combined at 0\,dB SNR (signals normalized to equal power). To ensure temporal overlap, the secondary sound is truncated or looped to match the primary’s duration. We then extract the RMS envelope of the primary and apply it to the mix, anchoring the composite to follow the macroscopic temporal ``behavior'' of the primary source. This produces a more realistic surrogate morph: rather than two sounds overlapping arbitrarily, the mix exhibits the temporal structure of the primary with added timbral detail from the secondary. 
\newline \textbf{\textit{Frequency-Domain Alignment (Spectral Interpolation)}}: Inspired by classic DSP interpolation, we employ a simple frequency-domain alignment trick to align the spectral content of two sounds. Given length-matched waveforms, we compute their FFT magnitudes and construct an averaged target spectrum: $|Y_{\text{target}}| = 0.5 \cdot|Y_1| +  0.5 \cdot|Y_2|$. This serves as a shared spectral envelope. For each sound, we derive a frequency-dependent gain mask (EQ curve) as the ratio between the target and original spectrum, specifying how much to boost or attenuate each band. These EQ curves are smoothed with a moving window, applied to the signals, and summed to yield the spectral composite. Perceptually, this produces a surrogate morph in which both sounds are projected into a common timbral space. 

\subsection{Augmentation Modes}
\textit{\textbf{Diverse Training Targets}:} To diversify training, we define four augmentation modes (Fig. \ref{fig:bigfig}): (i) RMS-only, (ii) Spectral-only, (iii) Both (Spectral+RMS), or (iv) None (unaugmented mix). During training, one mode is randomly assigned to each audio pair. \newline \textit{\textbf{Caption Conditioning}:} Each augmentation mode is paired with a corresponding caption: (i) RMS-only, where captions indicate primary behavior with textures from both \textit{(``The behavior of $X$ with textures from $X$ and $Y$'')}, (ii) Spectral-only, highlighting timbral blending \textit{(``A spectral blend of $X$ and $Y$'')}, (iii) Both, describing primary behavior with blended timbres \textit{(``The behavior of $X$ with a spectral blend of $X$ and $Y$'')}, and (iv) None \textit{(``A mix of $X$ and $Y$'')}.  

\section{Morphing Evaluation Metrics}
\label{sec:metrics}
A generative model can produce a morph, a mix, or collapse into a single-concept generation. To our knowledge, there is no standardized metric for distinguishing morphs from mixes or single-concept generations. To explore potential proxies, we tested the \textit{compressibility} of various low-level features (MFCCs, DAC \cite{kumar2023high} latents, or log-melspectrograms) via dimensionality reduction, hypothesizing that true morphs, and especially single-concept generations, form more coherent, lower-dimensional representations, while mixes retain more independent variability. To validate candidate proxies, we conducted a human study on 250 clips spanning mixes, single-concept generations, and morphs, annotated by a professional sound designer and the first author. 
We then compared features against these annotations using Spearman correlations and ROC–AUC scores. The most reliable indicator was the cumulative variance explained by the first two principal components of the DAC latent space, which we term the \textbf{Latent Compressibility Score (LCS)}. LCS correlated strongly with humans labeling audio as a morph (Spearman's $\rho$=0.9, p=0.037), 
suggesting it as a practical proxy for a system’s tendency to produce morphs. 

While LCS measures whether an output is perceived as a morph at all, the following metrics assess the semantic composition of that morph. To measure perceptual-relevance, we follow Caetano \& Osaka's \cite{caetano2012formal} morphing criteria of \textit{correspondence} and \textit{intermediateness}. We adopt an embedding-based approach, using FLAM \cite{wu2025flam}, a shared text–audio model, to compute clip-wise cosine similarities between the morphed audio and the text embeddings of primary and secondary source concepts, $sim_X$ and $sim_Y$ respectively. \textbf{Correspondence} is defined as the joint semantic presence of both source concepts in the morph. As this requires both similarities to be high, we operationalize this as the harmonic mean of $sim_X$ and $sim_Y$ (analogous to F1 scoring): $\tfrac{2\ \cdot  sim_X \cdot sim_Y}{sim_X + sim_Y}$.

\textbf{Intermediateness} captures balance between concepts; higher when the morph is equally similar to both concepts and lower when one dominates. We compute this as the normalized similarity difference, $1 - \tfrac{|sim_X - sim_Y|}{\max(sim_X, sim_Y)}$.

Due to the asymmetric nature of sound infusions, we extend evaluation with a softmax-based \textbf{directionality} score to assess a morph's alignment with the intended prompt direction \textit{(``behavior like X, timbre like Y'')} versus its reverse \textit{(``behavior like Y, timbre like X'')}. 
We compute cosine similarities between the audio embedding and the intended ($s_{\text{int}}$) and reversed ($s_{\text{rev}}$) prompt embeddings and apply a softmax ($T=0.05)$ rather than raw difference to emphasize relative directional preference, resolving small but consistent directional biases detectable under semantic overlap. The resulting probability $p$ is mapped to $[-1,1]$ via $2p-1$, where +1 indicates bias towards intended prompt and -1 towards the reverse. We also report FAD \cite{kilgour2018fr} as an audio quality metric with reference embeddings computed from a proprietary dataset of 40k high-quality SFX with a sound-class distribution matched to the training data. 

\section{Experiments \& Results}
\label{sec:experiments}

\subsection{Curating Concept Pairs for Evaluation}
\label{sec:conceptpairs}
We curated a set of 50 concept pairs of distinct sound classes designed to test morphing performance, informed by sound-design practices \cite{hillman2014craftsman, pinch2012oxford, chion2012three} and taxonomies\footnote{https://universalcategorysystem.com/}. The set aims for categorical coverage across source types (e.g., impact/temporally localized events, continuous textures) and infusion types, with an emphasis on \textit{inter-class} blends that are more difficult in practice to achieve (e.g., blending a TNT explosion with a choir swell) rather than intra-class blends (e.g., a soft whoosh with a sparky, metallic whoosh). Each pair is used in both directions (``behavior of $X$, timbre like $Y$'' \& vice versa), yielding 100 sound infusion prompts total.

\begin{table}[t]
\centering
\setlength{\tabcolsep}{3pt}
\resizebox{\columnwidth}{!}{%
\begin{tabular}{@{}p{2.0cm}|cc|cccc|ccccc@{}}
\toprule
Model 
& \rotatebox{90}{$t$ start} 
& \rotatebox{90}{$t$ end} 
& \rotatebox{90}{RMS} 
& \rotatebox{90}{Spectral} 
& \rotatebox{90}{Both} 
& \rotatebox{90}{None} 
& \rotatebox{90}{LCS $\uparrow$} 
& \rotatebox{90}{Correspond. $\uparrow$} 
& \rotatebox{90}{Intermediate. $\uparrow$} 
& \rotatebox{90}{Direct. $\uparrow$} 
& \rotatebox{90}{FAD $\downarrow$} 
\\
\midrule

base 
  & - & - & - & - & - & - 
  & 0.136 & 0.678 & 0.611 & 0.525 & 1.219 \\
\midrule

\multirow{4}{*}{\makecell{+Timestep \\ Allocation}}
  & 0 & 1 & \cmark & \xmarkgray & \xmarkgray & \xmarkgray 
  & 0.128 & 0.699 & 0.646 & 0.173 & 1.230 \\
  & 0.25 & 1 & \cmark & \xmarkgray & \xmarkgray & \xmarkgray 
  & \underline{0.143} & 0.705 & \underline{0.658} & 0.278 & 1.235 \\
  & 0.5 & 1 & \cmark & \xmarkgray & \xmarkgray & \xmarkgray 
  & \underline{0.141} & \underline{0.721} & \textbf{0.672} & 0.296 & \underline{1.221} \\
  & 0.75 & 1 & \cmark & \xmarkgray & \xmarkgray & \xmarkgray 
  & 0.134 & 0.717 & 0.653 & \underline{0.364} & 1.225 \\
\midrule

\multirow{3}{*}{\makecell{+Augmentation \\ Mode}}
  & 0.5 & 1 & \cmark & \xmarkgray & \cmark & \xmarkgray 
  & 0.135 & 0.700 & 0.623 & 0.363 & 1.226 \\
  & 0.5 & 1 & \cmark & \cmark & \cmark & \xmarkgray 
  & \textbf{0.150} & \textbf{0.725} & 0.648 & \textbf{0.436} & \textbf{1.220} \\
  & 0.5 & 1 & \cmark & \cmark & \cmark & \cmark 
  & \underline{0.143} & 0.712 & 0.650 & 0.349 & 1.222 \\
\midrule
\midrule

Simple Mixing 
  & - & - & - & - & - & - 
  & 0.132 & \textbf{0.758} & \textbf{0.690} & -9.25e-13 & 1.293 \\
LGrS
  & - & - & - & - & - & - 
  & 0.173 & 0.539 & 0.638 & -0.119 & \underline{1.290} \\
MorphFader
  & - & - & - & - & - & - 
  & 0.085 & 0.418 & 0.421 & -9.72e-13 & 1.430 \\
SoundMorpher 
  & - & - & - & - & - & - 
  & \textbf{0.242} & 0.591 & 0.641 & -9.64e-13 & 1.380 \\
\textbf{Mix2Morph} 
  & 0.5 & 1 & \cmark & \cmark & \cmark & \xmarkgray 
  & \underline{0.150} & \underline{0.725} & \underline{0.648} & \textbf{0.436} & \textbf{1.220} \\
\bottomrule
\end{tabular}}
\caption{Ablation results (above) and baseline comparisons (below). Best scores bolded, second-best underlined (per section).}
\label{tab:bigtableresult}
\vspace{-1.0em}
\end{table}

\subsection{Model Implementation Details}
Our base model is a large text-to-audio (TTA) latent diffusion transformer pretrained to perform single sound generation, similar to \cite{evans2024long, garcia2025sketch2sound}. The architecture consists of a VAE that compresses 48 kHz stereo audio into 256-dimensional latent sequences at 40 Hz. Generated latent sequences are decoded back to waveform via the VAE decoder. The training data for both pretraining and fine-tuning is a large set of proprietary, licensed SFX datasets and publicly available CC-licensed general audio corpora. After pretraining, we fine-tune for 50k steps using the generated surrogate morph dataset (\S\ref{sec:methods}). Training uses 8s audio segments mixed via augmentation techniques; we generate 3s sounds for evaluation.

\subsection{Effect of Training at Higher Timesteps}
The base model, when given the two-concept infusion prompts, often collapses to single-concept outputs. The metrics reflect this imbalance: high directionality (0.525) and compressibility (LCS = 0.136), but weaker correspondence (0.678), reflecting limited joint representation of both sources. To test our hypothesis that noisy surrogate morphs are best utilized at higher timesteps, we restrict the mix dataset to specific diffusion timestep ranges during training. Outside the allocated window, training proceeds as in the base model, i.e., the single-source reconstruction objective. We fine-tune models with allocations of $[0,1]$, $[0.25,1]$, $[0.5,1]$, $[0.75,1]$, using RMS-only augmentation for simplicity, and compare against the unmodified base model (Table \ref{tab:bigtableresult}). Shifting augmented mixes toward higher timesteps improves performance relative to base, \textbf{peaking with $t \in [0.5,1]$, which yields the best balance and all-around performance:} highest correspondence (0.721) and intermediateness (0.672) while maintaining strong directionality (0.296). Further constricting to $[0.75,1]$ increases directionality (0.364) but reduces intermediateness, indicating a trade-off between preserving semantic balance and strengthening prompt asymmetry. Qualitatively, this aligns with perceptual impressions: $[0.5,1]$ best incorporates both sources while retaining the primary sound’s quality, indicating blending, whereas $[0.75,1]$ loses shared concepts and reverts back toward base-model behavior. We therefore fix the allocation range to $[0.5,1]$ in subsequent experiments.

\subsection{Effect of Augmentation Modes}
Fixing the timestep range to $t \in [0.5,1]$, we compare the effect of applying different augmentation modes to the mix at varying probabilities. In addition to the RMS-only model (probability= 1.0), we evaluate (i) a balanced configuration with RMS-only (0.5) and Both (0.5), and (ii) a diversified three-way configuration with RMS-only (0.33), Spectral-only (0.33), and Both (0.33). Results in Table \ref{tab:bigtableresult} show that RMS-only improves upon the baseline, but both richer augmentation configurations outperform RMS-only, with the \textbf{three-way configuration delivering the strongest gains across nearly all metrics} (with slightly lower but still competitive intermediateness, reflecting a trade-off with correspondence and directionality). It also yields the lowest FAD score, indicating high-quality generations. Including ``None'' as a fourth branch reduces performance across metrics. Qualitatively, the three-way configuration produces the most natural and well-blended infusions; the two-way and four-way branches both yielding more mix-like outputs. We thus adopt this three-way augmentation model, trained at $t \in [0.5,1]$, as our final training setup for downstream evaluation, i.e. \textbf{Mix2Morph}.

\subsection{Baselines for Comparison}
We compare Mix2Morph against five baselines: (i) base model, (ii) simple waveform mixing, (iii) latent granular resynthesis (LGrS) \cite{tokui2025latent}, (iv) MorphFader \cite{kamath2025morphfader}, and (v) SoundMorpher \cite{niu2024soundmorpher}. For simple mixing, the primary and secondary sources are generated independently with the base model (without Mix2Morph finetuning) and then mixed. LGrS is included as it explicitly addresses sound infusion; we re-implement it using a 48 kHz VAE (the same as our base diffusion model) instead of the original 44.1 kHz DAC and provide it with the same input pairs as simple mixing. MorphFader requires paired text prompts (e.g., ``alien voice'' and ``lion roaring''), which we adapt for infusion-style comparisons, while SoundMorpher functions as an audio-to-audio method with only auxiliary text conditioning, so we provide it with the same input pairs as LGrS and mixing. We note that Mix2Morph, LGrS, and simple mixing operate at 48 kHz, whereas MorphFader and SoundMorpher are restricted to 16 kHz as an artifact of AudioLDM2.

\subsection{Quantitative Results}
\label{sec:objectivecompare}
Shown in Table \ref{tab:bigtableresult}, Mix2Morph consistently outperforms all baselines. It achieves higher correspondence and intermediateness than LGrS, MorphFader, and SoundMorpher while maintaining positive directionality, whereas the latter two collapse to near-zero or negative scores (though expected since they were not designed for infusion). Simple mixing attains high correspondence (0.758) but near-zero directionality, confirming it produces overlays rather than coherent morphs. Mix2Morph also yields higher LCS than the base model and mixing baselines (0.141 vs 0.127–0.136), indicating stronger morph tendencies, though slightly lower than LGrS and SoundMorpher, which is expected given their more aggressive latent operations. However, LGrS and SoundMorpher both tend to show lower correspondence, i.e., effectively producing morphs, but ones that are less perceptually coherent, whereas Mix2Morph achieves a better balance between morph strength and perceptual fidelity. Relative to the unmodified base model, Mix2Morph improves correspondence (0.721 vs 0.678) and intermediateness (0.672 vs 0.611), reduces directionality (0.296 vs 0.525), and raises LCS (0.141 vs 0.136), effectively reorienting the base model from its single-source bias toward more balanced blends that still preserve the primary sound’s identity. \textbf{Overall, Mix2Morph delivers the best balance of semantic coherence and perceptual relevance without sacrificing fidelity, enabling more coherent and perceptually balanced infusions than both the base model and prior baselines}.

\begin{figure}[t]
\centering
\includegraphics[width=0.49\textwidth]{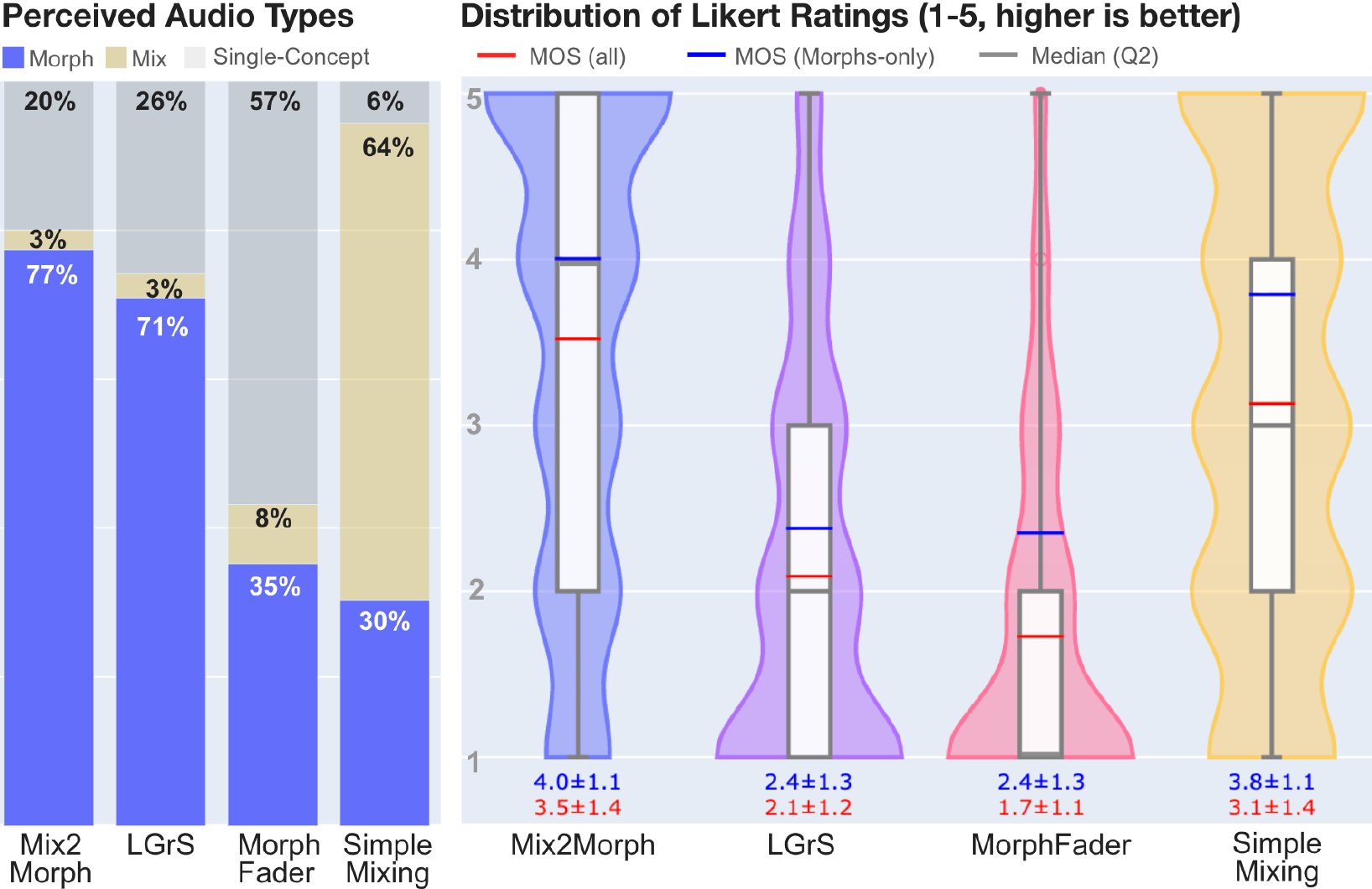}
\caption{Listener Study Results: Mix2Morph achieves the highest morph rate, with MOS skewed toward higher values.}
\label{fig:listening_results_basic}
\vspace{-1.0em}
\end{figure}

\subsection{Subjective Evaluation}
We conducted a listening test with $N=25$ participants (inc. audio producers, musicians, researchers, content creators, and enthusiasts). 
From the 100 infusion prompts in \S\ref{sec:conceptpairs}, we randomly select 20. 
To mitigate rater fatigue, we limit evaluation to three external baselines: simple mixing, LGrS \cite{tokui2025latent}, and MorphFader \cite{kamath2025morphfader} (excluding SoundMorpher due to redundancy with AudioLDM2 and high computational cost).

Given a generated sound infusion and its target infusion prompt (e.g., \textit{``behavior of a dog barking with timbre like a car horn'')}, we asked participants to score each generated output on this question: \textit{How successfully does the audio capture the morphing prompt above?} using a 1-5 Likert scale, where 5 is best. Participants were also asked to label each generated output as one of the following: \textit{a Morph, a Mix, or Single-Concept}. Each participant evaluated 80 morphs (4 models $\times$ 20 prompts), randomized per subject. 

We report audio type labels and Mean Opinion Score (MOS) in Fig. \ref{fig:listening_results_basic}. A repeated-measures ANOVA showed a significant main effect of model on MOS ($F(3,72)=76.4, p<0.001$). Tukey tests confirmed that Mix2Morph scored significantly higher than LGrS and MorphFader (both $p<0.001$). The morph rate is the percentage of generated outputs labeled \textit{Morph}.\textbf{ Mix2Morph achieved the highest morph rate (77\%) and MOS (3.52 overall; 4.00 morph-MOS)}, 
consistent with its strong objective metric scores (\S\ref{sec:objectivecompare}). LGrS produced a similar morph rate (71\%) but much lower quality (MOS = 2.09), reflecting weaker perceptual coherence despite decent correspondence in objective scoring. Simple mixing was often judged as mixes (64\%) but still reached moderate MOS (3.13 overall; 3.79 morph-MOS), suggesting it can suffice for certain infusion types (e.g., impact sounds with aligned onsets). Consistent with objective scores, MorphFader reached a morph rate of 35\%, producing more single-concept and mix-like outputs, and the lowest MOS (1.73), likely due to its 16kHz backbone. Overall, Mix2Morph outperformed all baselines in MOS and morph frequency. 

\section{Conclusion}
\label{sec:Conclusion}
We introduce Mix2Morph, a TTA model that leverages surrogate mixes to adapt pretrained audio diffusion models for morphing without a dedicated morph dataset. By leveraging augmented mixes as noisy training data at higher diffusion timesteps, Mix2Morph achieves consistent improvements in morphing quality across both objective and subjective evaluations, outperforming existing baselines. This represents a substantial step toward more controllable and concept-driven tools for sound design. Future work may explore more intuitive control mechanisms (e.g., audio-to-audio).
\vfill\pagebreak

\bibliographystyle{IEEEbib}
\bibliography{refs}

\begin{thebibliography}{10}

\bibitem{caetano2011morphing}
Marcelo Caetano,
\newblock {\em Morphing isolated quasi-harmonic acoustic musical instrument sounds guided by perceptually motivated features},
\newblock Ph.D. thesis, Paris 6, 2011.

\bibitem{heller2022hybrid}
Laurie~M Heller and Lauren Wolf,
\newblock ``When hybrid sound effects are better than real recordings,''
\newblock in {\em Proceedings of Meetings on Acoustics}. Acoustical Society of America, 2022, vol.~46, p. 050002.

\bibitem{oxenham2018we}
Andrew~J Oxenham,
\newblock ``How we hear: The perception and neural coding of sound,''
\newblock {\em Annual review of psychology}, vol. 69, no. 1, pp. 27--50, 2018.

\bibitem{smith2002chimaeric}
Zachary~M Smith, Bertrand Delgutte, and Andrew~J Oxenham,
\newblock ``Chimaeric sounds reveal dichotomies in auditory perception,''
\newblock {\em Nature}, vol. 416, no. 6876, pp. 87--90, 2002.

\bibitem{Iverson1993IsolatingTDA}
P.~Iverson and C.~Krumhansl,
\newblock ``Isolating the dynamic attributes of musical timbre.,''
\newblock {\em The Journal of the Acoustical Society of America}, vol. 94 5, pp. 2595--603, 1993.

\bibitem{dixit2025learning}
Satvik Dixit, Sungjoon Park, Chris Donahue, and Laurie~M Heller,
\newblock ``Learning perceptually relevant temporal envelope morphing,''
\newblock in {\em Proceedings of the IEEE Workshop on Applications of Signal Processing to Audio and Acoustics (WASPAA)}, 2025.

\bibitem{niu2024soundmorpher}
Xinlei Niu, Jing Zhang, and Charles~Patrick Martin,
\newblock ``Soundmorpher: Perceptually-uniform sound morphing with diffusion model,''
\newblock {\em arXiv preprint arXiv:2410.02144}, 2024.

\bibitem{sethares2015kernel}
William~A Sethares and James~A Bucklew,
\newblock ``Kernel techniques for generalized audio crossfades,''
\newblock {\em Cogent Mathematics}, vol. 2, no. 1, pp. 1102116, 2015.

\bibitem{caetano2012formal}
Marcelo Caetano and Naotoshi Osaka,
\newblock ``A formal evaluation framework for sound morphing,''
\newblock in {\em ICMC}, 2012.

\bibitem{tokui2025latent}
Nao Tokui and Tom Baker,
\newblock ``Latent granular resynthesis using neural audio codecs,''
\newblock {\em arXiv preprint arXiv:2507.19202}, 2025.

\bibitem{kazazis2016sound}
Savvas Kazazis, Philippe Depalle, and Stephen McAdams,
\newblock ``Sound morphing by audio descriptors and parameter interpolation,''
\newblock in {\em Proceedings of the 19th International Conference on Digital Audio Effects (DAFx-16). Brno, Czech Republic}, 2016.

\bibitem{kamath2025morphfader}
Purnima Kamath, Chitralekha Gupta, and Suranga Nanayakkara,
\newblock ``Morphfader: Enabling fine-grained controllable morphing with text-to-audio models,''
\newblock in {\em ICASSP 2025-2025 IEEE International Conference on Acoustics, Speech and Signal Processing (ICASSP)}. IEEE, 2025, pp. 1--5.

\bibitem{caetano2019morphing}
Marcelo Caetano,
\newblock ``Morphing musical instrument sounds with the sinusoidal model in the sound morphing toolbox,''
\newblock in {\em International Symposium on Computer Music Multidisciplinary Research}. Springer, 2019, pp. 481--503.

\bibitem{donahue2016extended}
Chris Donahue, Tom Erbe, and Miller~S Puckette,
\newblock ``Extended convolution techniques for cross-synthesis,''
\newblock in {\em ICMC}, 2016.

\bibitem{Puche2021CaesynthRT}
Aaron~Valero Puche and Sukhan Lee,
\newblock ``Caesynth: Real-time timbre interpolation and pitch control with conditional autoencoders,''
\newblock {\em 2021 IEEE 31st International Workshop on Machine Learning for Signal Processing (MLSP)}, pp. 1--6, 2021.

\bibitem{gupta2023towards}
Chitralekha Gupta, Purnima Kamath, Yize Wei, Zhuoyao Li, Suranga Nanayakkara, and Lonce Wyse,
\newblock ``Towards controllable audio texture morphing,''
\newblock in {\em ICASSP 2023-2023 IEEE International Conference on Acoustics, Speech and Signal Processing (ICASSP)}. IEEE, 2023, pp. 1--5.

\bibitem{liu2024audioldm}
Haohe Liu, Yi~Yuan, Xubo Liu, Xinhao Mei, Qiuqiang Kong, Qiao Tian, Yuping Wang, Wenwu Wang, Yuxuan Wang, and Mark~D Plumbley,
\newblock ``Audioldm 2: Learning holistic audio generation with self-supervised pretraining,''
\newblock {\em IEEE/ACM Transactions on Audio, Speech, and Language Processing}, vol. 32, pp. 2871--2883, 2024.

\bibitem{bitton2020neural}
Adrien Bitton, Philippe Esling, and Tatsuya Harada,
\newblock ``Neural granular sound synthesis,''
\newblock in {\em Proceedings of the International Computer Music Conference (ICMC)}, 2020.

\bibitem{tralie2024concatenator}
Christopher~J. Tralie and Ben Cantil,
\newblock ``The concatenator: A bayesian approach to real time concatenative musaicing,''
\newblock in {\em Proceedings of the 25th Conference of the International Society for Music Information Retrieval (ISMIR)}, 2024.

\bibitem{daras2025ambient}
Giannis Daras, Adrian Rodriguez-Munoz, Adam Klivans, Antonio Torralba, and Constantinos Daskalakis,
\newblock ``Ambient diffusion omni: Training good models with bad data,''
\newblock {\em arXiv preprint arXiv:2506.10038}, 2025.

\bibitem{kumar2023high}
Rithesh Kumar, Prem Seetharaman, Alejandro Luebs, Ishaan Kumar, and Kundan Kumar,
\newblock ``High-fidelity audio compression with improved rvqgan,''
\newblock {\em Advances in Neural Information Processing Systems}, vol. 36, pp. 27980--27993, 2023.

\bibitem{wu2025flam}
Yusong Wu, Christos Tsirigotis, Ke~Chen, Cheng-Zhi~Anna Huang, Aaron Courville, Oriol Nieto, Prem Seetharaman, and Justin Salamon,
\newblock ``{FLAM}: Frame-wise language-audio modeling,''
\newblock in {\em Forty-second International Conference on Machine Learning}, 2025.

\bibitem{kilgour2018fr}
Kevin Kilgour, Mauricio Zuluaga, Dominik Roblek, and Matthew Sharifi,
\newblock ``Fréchet audio distance: A reference-free metric for evaluating music enhancement algorithms,''
\newblock in {\em Interspeech 2019}, 2019, pp. 2350--2354.

\bibitem{hillman2014craftsman}
Neil Hillman and Sandra Pauletto,
\newblock ``The craftsman: The use of sound design to elicit emotions,''
\newblock {\em The Soundtrack}, vol. 7, no. 1, pp. 5--23, 2014.

\bibitem{pinch2012oxford}
Trevor Pinch and Karin Bijsterveld,
\newblock {\em The Oxford handbook of sound studies},
\newblock OUP USA, 2012.

\bibitem{chion2012three}
Michel Chion,
\newblock ``The three listening modes,''
\newblock {\em The sound studies reader}, vol. 1, 2012.

\bibitem{evans2024long}
Zach Evans, Julian Parker, CJ~Carr, Zack Zukowski, Josiah Taylor, and Jordi Pons,
\newblock ``Long-form music generation with latent diffusion,''
\newblock in {\em International Society for Music Information Retrieval Conference}, 2024.

\bibitem{garcia2025sketch2sound}
Hugo~Flores Garc{\'\i}a, Oriol Nieto, Justin Salamon, Bryan Pardo, and Prem Seetharaman,
\newblock ``Sketch2sound: Controllable audio generation via time-varying signals and sonic imitations,''
\newblock in {\em ICASSP 2025-2025 IEEE International Conference on Acoustics, Speech and Signal Processing (ICASSP)}. IEEE, 2025, pp. 1--5.

\end{thebibliography}



\end{document}